\documentclass[aps,showpacs,twocolumn,superscriptaddress,nofootinbib,floatfix,section 10pt]{revtex4-1}
\usepackage{amsfonts,amssymb,stmaryrd,latexsym,amsmath,braket}
\usepackage{graphicx,subfigure}
\usepackage{comment}
\usepackage{times}
\usepackage{slashed}
\usepackage{bm}
\usepackage{braket}
\usepackage{appendix}
\usepackage{enumitem}
\usepackage{multirow}
\usepackage{array}

\usepackage[colorlinks=true,backref=true,linktocpage=true,citecolor=blue,urlcolor=blue,linkcolor=blue,pdfpagemode=UseOutlines]{hyperref}
\usepackage[dvipsnames]{xcolor}

\newcommand{\slfrac}[2]{\left.#1\middle/#2\right.}

\begin{document}
\title{Optimizing rodeo projection}
\author{Thomas D. Cohen}
\email{cohen@umd.edu}
\affiliation{Department of Physics and Maryland Center for Fundamental Physics, University of Maryland, College Park, MD 20742 USA}

\author{Hyunwoo Oh}
\email{hyunwooh@umd.edu}
\affiliation{Department of Physics and Maryland Center for Fundamental Physics, University of Maryland, College Park, MD 20742 USA}

\begin{abstract}
The rodeo algorithm has been proposed recently as an efficient method in quantum computing for projection of a given initial state onto a state of fixed energy for systems with discrete spectra. In the initial formulation of the rodeo algorithm, evolution times were chosen randomly via a Gaussian distribution with a fixed standard deviation. In this paper, it is shown that such a random approach for choosing times suffers from exponentially large fluctuations in the suppression of unwanted components: as the number of iterations gets large, the distribution of suppression factors obtained from random selection approaches a log-normal distribution leading to remarkably large fluctuations. We note that by choosing times intentionally rather than randomly, such fluctuations can be avoided and strict upper bounds on the suppression can be obtained. Moreover, the average suppression using fixed computational cost can be reduced by many orders of magnitude relative to the random algorithm. A key to doing this is to choose times that vary over exponentially many time scales, starting from a modest maximum scale and going down to time scales exponentially smaller.
\end{abstract}

\date{\today}

\maketitle

\section{Introduction}

Quantum computers, suggested by Richard Feynman in the 1980s~\cite{Feynman:1981tf}, are expected to solve problems that are not accessible to classical computers. It is known that for a number of problems the asymptotic scaling of quantum algorithms are better than that of the best-known algorithms with classical computers. A notable example is the factoring of large numbers, where Shor's algorithm~\cite{Shor:1994jg} has a computational cost that is a polynomial in the logarithm of the size of the system while the best-known classical algorithms have a cost that grows faster than any power law~\cite{10.2307/1971363, 1570009750725490304, pomerance1996tale, Brent1980}. It is widely believed that many quantum mechanical problems, which suffer from exponentially bad sign problems~\cite{Troyer:2004ge} when computed classically via Monte Carlo methods, would only require polynomial resources on a quantum computer. Examples include real-time dynamics~\cite{10.1063/1.445732, Alexandru:2016gsd}, where the time evolution induces complex phases that induce sign problems, and certain problems in which a chemical potential for fermions, such as QCD at a finite chemical potential~\cite{Aarts:2015tyj}, induces an complex effective action after integrating out the fermions. This latter example is of great importance in nuclear physics since QCD behaves this way. Thus, quantum computing is potentially useful for simulating a wide range of quantum systems with exponentially lower cost, spanning fields  from condensed matter physics to nuclear physics and particle physics~\cite{Jordan:2012xnu, Zohar:2014qma, Georgescu:2013oza, Bauer:2022hpo, Funcke:2023jbq, Beck:2023xhh}. 

To simulate quantum physics, one needs to prepare initial states and evolve them in time according to the underlying hamiltonian. It has long been recognized that the state preparation problem is a key to viable quantum computation and research over nearly three decades has focused on developing methods for state preparation with a focus on using minimal resources~\cite{Kitaev:1995qy, 10.1063/1.3115177, Abrams:1997gk, PhysRevA.65.042323, https://doi.org/10.1002/qua.26853, Bharti:2021zez, PhysRevLett.129.230504, PhysRevApplied.15.034027, Lee:2019zze, PhysRevA.78.042307, Kaplan:2017ccd, Gustafson:2020vqg}.

The rodeo algorithm~\cite{Choi:2020pdg} has been suggested as a cost-effective method of preparing quantum states by projecting a given initial state onto an energy eigenstate for systems with discrete spectra; the success rate is proportional to the square of the overlap between the initial state and the target state of interest and the time needed is logarithmic in the accuracy making it exponentially faster than phase estimation algorithm. It has been performed on a superconducting quantum computer~\cite{Qian:2021wya} and variations of the algorithm have been considered~\cite{Bee-Lindgren:2022nqb}. An alternative approach, also based on a projection method via filtering processes has been suggested~\cite{Stetcu:2022nhy}, which is designed to exploit particular knowledge about the character of the state.

The rodeo algorithm can be used either as a direct method of state preparation or in conjunction with other techniques for state preparation such as adiabatic quantum computing~\cite{farhi2000quantum, Wan:2020fwj, aharonov2003adiabatic, CoelloPerez:2021jkh, Ciavarella:2022skg, RevModPhys.90.015002, Hauke_2020}---which could act to extend its utility in state preparation to situations in which the overlap between the initial and final states is small.

The rodeo algorithm is based on quantum interferometry and needs to be done for a relatively large number of iterations to ensure high accuracy; each iteration is associated with a running time and accurate projections require numerous distinct times. It is called the rodeo algorithm since during each iteration there is a probability that states other than the desired state are thrown out of the state in analogy to a hapless bull rider in a rodeo.

In Ref.~\cite{Choi:2020pdg}, the emphasis was on the suppression factor per iteration. This might create an impression that the optimal approach is to create the maximum suppression per iteration and then perform a sufficient number of iterations to achieve the desired suppression. However, this need not be the case: large suppression factors per iteration require large times. It can be more efficient in terms of total time to use more iterations with shorter times per iteration (and less suppression per iteration). 

The goal of this paper is to study how to optimize the algorithm to obtain the most accurate version of the algorithm with minimal computational costs. As the dominant computational cost is running a Hamiltonian over time, we will use the net time expended as a proxy for computational costs. The algorithm involves multiple iterations of a basic computational structure, which is run for a given amount of time; optimization involves deciding what times should be used for each iteration and how many iterations should be run.

In this context, we note that if the computation is to be done on a digital quantum computer based on gates, then ``time'' could represent a mathematical time in a Trotterized version~\cite{Trotter:1959, Suzuki:1976be} of the time evolution. However, the rodeo algorithm would also be viable on analog quantum simulators---provided that one could use quantum gates to turn the simulator on and off in a controlled way. It is not implausible that such a scheme of gate-controlled analog quantum-simulation (GCAQS) might be the first setting in which rodeo projection proves to be decisive.

While the rodeo algorithm can be used to project onto any discrete state of the spectrum for any Hermitian operator, for simplicity of discussion, this paper will focus on the projection onto the ground state of a Hamiltonian of physical interest---which is likely to be a common application. Moreover, to further simplify discussion, we will assume that a constant has been added to the Hamiltonian of the system to make the ground state energy zero. However, the discussion given here goes through {\it mutatis mutandis} for the more general case of an arbitrary discrete state with the any value of the ground state energy.

In this paper, we demonstrate that the version of the rodeo algorithm shown in Ref.~\cite{Choi:2020pdg}, which is based on choosing the evolution times for the iterations randomly using a Gaussian distribution with a fixed standard deviation, has large fluctuations in efficiency that grow exponentially with the number of iterations. We show that these fluctuations can be suppressed by a very efficient algorithm designed with a set of ``super iterations'' which include iterations of multiple time scales.
 
We begin in Section~\ref{sec2} with an introduction to the rodeo projection algorithm; additional details are given in Appendix \ref{sec:it}. Section~\ref{OptRRA} describes how the rodeo algorithm whose times for each iteration are chosen randomly via a Gaussian distribution can be optimized to minimize computational costs while maximizing suppression of unwanted states. This will serve as a baseline for comparison with other approaches. It is shown in Section~\ref{sec4} that this random version of the rodeo algorithm has large fluctuations, which grow exponentially in the number of iteration. This implies that if one wishes to ensure a fixed level of accuracy, with a high level of confidence one would need to use substantially more resources than predicted by the average value. In order to suppress these fluctuations, in Section~\ref{Super}, the notion of a ``super iteration'' algorithm is introduced. In Section~\ref{sec6}, a simple {\it ad hoc} prescription of choosing super iterations is considered; additional details are given in Appendix \ref{App:WAM}. Its results are compared with the random approach and shown to yield substantially stronger suppression of unwanted components than the random approach. Finally, a few practical issues associated with the implementation of the rodeo algorithm are considered in Section~\ref{sec7}.

\section{Rodeo projection}\label{sec2}
The rodeo projection algorithm acts by suppressing components of the initial state depending on the energy of the state. Since the algorithm is only applicable to discrete states, there is necessarily an excited state with a minimum excitation, which we denote $\Delta$. It is extremely helpful to know the value of $\Delta$ when attempting to optimize the performance of the algorithm: clearly it makes no sense to spend computational resources in order to improve the suppression of would-be components with energies less than $\Delta$ as such components do not exist.

The existence of a minimum excitation energy $\Delta$, defines a natural time scale for our problem:
\begin{equation}
T_0 = \frac{2 \pi \hbar}{\Delta}
\end{equation}
which is the time needed for a full period for the phase evolution of the lowest excited state.

Using time as our proxy for computational cost is slightly problematic in that it is a dimensionful quantity. One can rescale all times in the problem by a constant factor and rescale all energies by its inverse and so the computational difficulty is the same. Moreover, a large extent in the analysis of the rodeo algorithm involves products of energies and times. Accordingly we will often express quantities in terms of the following dimensionless combinations:
\begin{equation}
\zeta = \frac{ E T}{2 \pi \hbar }=\frac{E}{\Delta} \frac{T}{T_0},   \; \; \; \zeta_{\rm tot} = \frac{E T_{\rm tot}}{2 \pi \hbar}=\frac{E}{\Delta} \frac{T_{\rm tot}}{T_0},
\label{Eq:zeta} \end{equation}
where $T$ is a time that occurs during one iteration and $T_{\rm tot}$ is the total time. The factor of $2 \pi$ is included so that $\zeta$ measures the number of periods.

The rodeo algorithm involves dynamics in which the physical system of interest is entangled with an ancilla qubit. The scheme is based on iterations, with the concept of ``successful'' and ``unsuccessful'' iterations determined by a measurement of the ancilla qubit, indicating whether an iteration is successful or not. The input to the $j^{\rm th}$ iteration is the physical state $|\psi \rangle_{\rm phys}^{j \, \rm initial}$ and the  output of successful iterations is the physical state $|\psi \rangle_{\rm phys}^{j \, \rm final}$; the time used to implement the iteration is $T_j$.  

Consider an initial physical state prior to the $1^{\rm st}$ iteration given by
\begin{equation}
\begin{split}
    |\psi\rangle_{\rm phys}^{\rm initial} & =  \alpha_g |\psi_g \rangle + \sum_c  \alpha_c|\psi_c \rangle \; \; {\rm with } \\  P_g^i & =|\alpha_g|^2 \; \;  {\rm and} \; \; |\alpha_g|^2+\sum_c  |\alpha_c|^2=1  \; \; \label{Eq:initialstate}
    \end{split}
\end{equation}
where the subscript $g$ indicates the ground state component and $c$ indicates the excited states (where $\hat{H}_{\rm phys} |\psi_a\rangle = E_a |\psi_a\rangle$); $P_g^i$ is the initial probability that the system is in the ground state.  The central features of the algorithm are
 
\begin{enumerate}[label=\roman*.]
    \item Following $n$ consecutive successful iterations, the $c^{\rm th}$ excited state is suppressed compared to the ground state by a factor of 
    \begin{equation}
        \begin{split}
    s_c &= \prod_{j=1}^n s_c^j \; \; {\rm with } \\ 
    s_c^j & = \cos^2 \left(\frac{\omega_c T_j}{2} \right) =   \cos^2 \left(\pi \zeta_c^j \right) \; \;  {\rm where } \\    \omega_c & \equiv \frac{E_c}{\hbar} \; \;  {\rm and} \; \;  \zeta_c^j \equiv \frac{\omega_c T^j}{2 \pi} .  \end{split}\label{Eq:suppress}\end{equation}
    One can represent the overall suppression factor after $n$ iterations for an arbitrary component in terms of $\zeta_{\rm tot}$ (given in Eq.~(\ref{Eq:zeta})) as 
    \begin{equation} \begin{split}
    s(\zeta_{\rm tot}) &=   s\left(\frac{E_c}{\Delta} \frac{T_{\rm tot}}{T_0} \right ) =  \prod_{j=1}^n  \cos^2 \left(\frac{\omega_c T_j}{2} \right)     \\    {\rm where} \; \; &T_{\rm tot}  =\sum_{k=1}^n T_n \; .
        \end{split}
    \end{equation}
    \label{f1}
    
    \item  A consequence is that $P^{n}_g$, the probability that the physical system is in the ground state after $n$ consecutive successful iterations, is
    \begin{subequations}
    \begin{equation}  
    \begin{split}
        P^n_g = & \frac{|\alpha_g|^2}{|\alpha_g|^2 + \sum_c \left(\prod_{j=1}^n  \cos^2 \left(\pi \zeta_c^j \right) \right) |\alpha_c|^2} \\
        = & \frac{P_g^i}{P_g^i + \int_{\Delta}^{\infty} d E\, \,  s\left(\frac{E}{\Delta} \frac{T_{\rm tot}}{T_0} \right ) \rho(E)}  \\
        =&  \frac{P_g^i}{P_g^i + (1-P_g^i)S_E } \label{Eq:Pg}
    \end{split} \end{equation} 
    where $s\left(\frac{E}{\Delta} \frac{T_{\rm tot}}{T_0} \right )$  was given above.  It is useful to consider the overall suppression for all excited states.  This is naturally given in terms of the spectral density of the initial state,  $\rho(E)$,  
    \begin{equation}
    \rho(E) \equiv {}^{\rm initial}_{\; \;\rm phys}\langle\psi|\delta(E-\hat{H})|\psi\rangle_{\rm phys}^{\rm initial}. \label{Eq:SpecDen}
    \end{equation}
    $S_E$, the overall suppression factor for excited states, is then given by 
    \begin{equation}
    \begin{split}
        S_E  \equiv &\frac{ \int_{\Delta-\epsilon}^{\infty} d E\, \,  s\left(\frac{E}{\Delta} \frac{T_{\rm tot}}{T_0} \right ) \rho(E)}{ \int_{\Delta-\epsilon}^{\infty}  dE\, \rho(E)} \,  
        {\rm with} \; \; 0 < \epsilon <\Delta. \; 
    \end{split}\label{Eq:SE}
    \end{equation}
    \end{subequations}
    (The integral starts at $\Delta - \epsilon$ to ensure that the delta function for the first (discrete) excited state is included.)  Clearly, $S_E$ goes to zero as  $n \rightarrow \infty$, which implies that as $n \rightarrow \infty$,  $P_g^{n} \rightarrow 1$. \label{f2}

    \item Assuming the ground state energy is known with very high accuracy and the algorithm is implemented without errors, the probability that $n$ consecutive successful iterations occur starting from    $|\psi \rangle_{\rm phys}^{j \, \rm initial}$, $ P^{n \, {\rm s}}$, is given by
    \begin{equation}
        P^{n \, \rm s} =  |\alpha_g|^2 + \sum_c \left(\prod_{j=1}^n  \cos^2 \left(\pi \zeta_c^j \right) \right) |\alpha_c|^2
    \end{equation}
    where $|\psi \rangle_{\rm phys}$ is the physical state after $n$ consecutive successful iterations. This implies that 
    \begin{equation} 
        \lim_{n\rightarrow \infty}  P^{n \, \rm s} \rightarrow  |\alpha_g|^2 =|\langle \psi_g |\psi \rangle_{\rm phys}^{\rm initial}|^2.
    \end{equation} 
    \label{f3}
    
    \item If an iteration is unsuccessful prior to obtaining a physical state with sufficiently high probability to be in the ground state (for the purposes of the problem one is studying), one needs to recreate the initial state and start from the beginning.  The expected number of times one needs to create (or recreate) $|\psi \rangle_{\rm phys}^{\rm initial}$ is $\frac{1}{  P^{n \, \rm s}}$ where $n$ is the number of successful iteration needed to achieve the desired accuracy. \label{f4}
    
    \item The expected net time to obtain a successful projected state with the accuracy associated with $n$ iterations 
    \begin{equation}  
        T^{\rm expected} = \frac{T^{n \, {\rm s}}}{|\langle \psi_g |\psi \rangle_{\rm phys}^{\rm initial}|^2 } \label{Eq:total_time}
    \end{equation}
    where $T^{n \, {\rm s}}$ is the time needed for $n$ successful iterations; in Eq.~(\ref{Eq:total_time}) it is assumed that $ \sum_c \left(\prod_{j=1}^n \cos^2 \left(\pi \zeta_c^j\right) \right) |\alpha_c|^2 $ is negligibly small.
    \label{f5}
\end{enumerate}

To see how these features come about, it is useful to focus on the details of how a single iteration works; this is given in Appendix~\ref{sec:it}.  Prior to the beginning of the $j^{\rm th}$ iteration, the the physical system is in pure state given in the energy eigenbasis  by 
\begin{equation}
\begin{split}
|\psi \rangle_{\rm phys}^{j \, \rm initial} &= \alpha^j_g |\psi_g \rangle + \sum_c  \alpha^j_c|\psi_c \rangle  \\ {\rm with } & \; \;  |\alpha^j_g|^2+\sum_c  |\alpha^j_c|^2=1  \; . \label{Eq:initialstate}
\end{split}
\end{equation}
Thus, $ P^{j \, i }_{g}$, the initial probability for the system to be in the ground state at the onset of iteration $j$, is given by $P^{j \, i }_{g} = |\alpha^j_g|^2 $.
 
As shown in Appendix~\ref{sec:it} the effect of a successful iteration on the state is to affect the following transformation
\begin{subequations}
\begin{equation}
\begin{split}
|\psi\rangle_{\rm phys}^{j \, \rm initial} &\xrightarrow[{\rm iteration}]{{\rm successful}} |\psi\rangle_{\rm phys}^{j \,  \rm final} \; \; \;  {\rm with} \\
|\psi \rangle_{\rm phys}^{j \, \rm final} &= \frac{ \alpha_g |\psi_g \rangle + \sum_c \frac{1}{2} \left( 1 + e^{-i 2 \pi \zeta_c^j}  \right) \alpha_c^j |\psi_c\rangle}{ \sqrt{ |\alpha_g^j|^2   + \sum_c |\alpha_c^j|^2 \cos^2\left(\pi \zeta_c^j \right) } }  \\
{\rm where} &\; \; \zeta_c^j  \equiv \frac{ E_c T^j}{2 \pi \hbar} \; .
\end{split}
\end{equation} 
Thus
\begin{equation}
\begin{split}
    |\psi \rangle_{\rm phys}^{j \, \rm final} &=\alpha'^j_g |\psi_g \rangle + \sum_c  \alpha'^j_c |\psi_c \rangle  \\ 
    {\rm with } \; \;   \alpha'^j_g &= \frac{\alpha^j_g} {\sqrt{ |\alpha_g^j|^2   + \sum_c |\alpha_c^j|^2 \cos^2\left(\pi \zeta_c^j \right) }} \label{Eq:newalph} \\
    \alpha'^j_c &= \frac{\alpha^j_c \cos\left(\pi \zeta_c^j \right) e^{-i \pi \zeta_c^j} } {\sqrt{ |\alpha_g^j|^2   + \sum_c |\alpha_c^j|^2 \cos^2\left(\pi \zeta_c^j \right) }} \; . \\
\end{split}
\end{equation}
\end{subequations}
A successful iteration of the algorithm increases the probability that the system is in the ground state:
\begin{equation}
P_{g }^{j \, 1  \, {\rm s} }   = |\alpha'^j_g|^2 = \frac{P_{g}^{j \,  i }}{ |\alpha_g^j|^2   + \sum_c |\alpha_c^j|^2 \cos^2\left(\pi \zeta_c^j \right) } \ge P_{g}^{j \,  i }
\label{Eq:PGC}
\end{equation}
where the fact that $|\alpha_g^j|^2 + \sum_c |\alpha_c^j|^2 \cos^2\left(\pi \zeta_c^j \right)$ is necessarily less than or equal to unity gives rise to the inequality.  Moreover this inequality is only saturated if every component $c$ either has $\alpha_c^j=0$ or has $\cos^2\left(\pi \zeta_c^j \right)=1$, an exceedingly unlikely possibility.  Thus a successful iteration will enhance the ground state component of the physical system.

From Eq.~(\ref{Eq:newalph}), it is clear that the enhancement of the ground state probability comes through the relative suppression of components other than the ground state, which is given by
\begin{equation}
    s_c^j\equiv \left( \slfrac{ \frac{|\alpha'^j_c|}{|\alpha'^j_g|} }{ \frac{|\alpha^j_c|}{|\alpha^j_g|} } \right)^2= \cos^2 \left(\pi \zeta_c^j \right) \; . \label{Eq:suppressj}
\end{equation}
Since this suppression happens for each iteration, the net suppression for component $c$ is the product $\prod_{j=1}^n s_c^j $  which establishes feature~\ref{f1} above; feature~\ref{f2} follows logically from it.

The reduced density matrix for the physical system is given in Eq.~(\ref{Eq:redden}) of Appendix \ref{sec:it}.  Since the reduced density matrix sums over the auxiliary qubit, it sums over both successful and unsuccessful iterations. From the form of Eq.~(\ref{Eq:redden}), it is apparent that the total probability that the physical system is in the ground state---including both successful and unsuccessful iterations---is given by  $|a_g^j|^2$, which as noted above is the initial probability of the system being in the ground state at the start of the $j^{\rm th}$ iteration. Thus, if one includes the contributions of both successful and unsuccessful iterations, the total probability that the system is in the ground state is unchanged by the iteration. Since this holds for all iterations, it follows that the net probability that the system is in the ground state is unchanged from the initial configuration prior to any iterations. Feature \ref{f3} follows immediately; features \ref{f4} and \ref{f5} are natural consequences of \ref{f3}

Once the number of iterations and the times for each iteration are fixed, the total suppression for components with any given energy is determined.  However, the algorithm as described so far has not been fully specified, since  criteria for choosing the times for each iteration and the number of iterations have not been given.  

It is useful to recall that the suppression factor for any iteration only depend on $\zeta_c^j =  E_c T^j/(2 \pi \hbar)$. Since our interest is to suppress contributions with various energies contained in the initial state, the goal is to pick times that usefully suppress these. Clearly, it is sub-optimal to use the same time for each iteration of algorithm: if this were done energies for which $w_c T^j$ is an odd-integer (where $\omega=E/\hbar$) would have a suppression of unity ({\it i.e} no suppression) no matter how many iterations were done and energies near these points would only be weakly suppressed. The remainder of the paper is dedicated to considering various schemes for picking the times.

\,

\section{Optimizing the random rodeo algorithm\label{OptRRA}}

\begin{figure*}[t!]
    \includegraphics[width=0.33\textwidth]{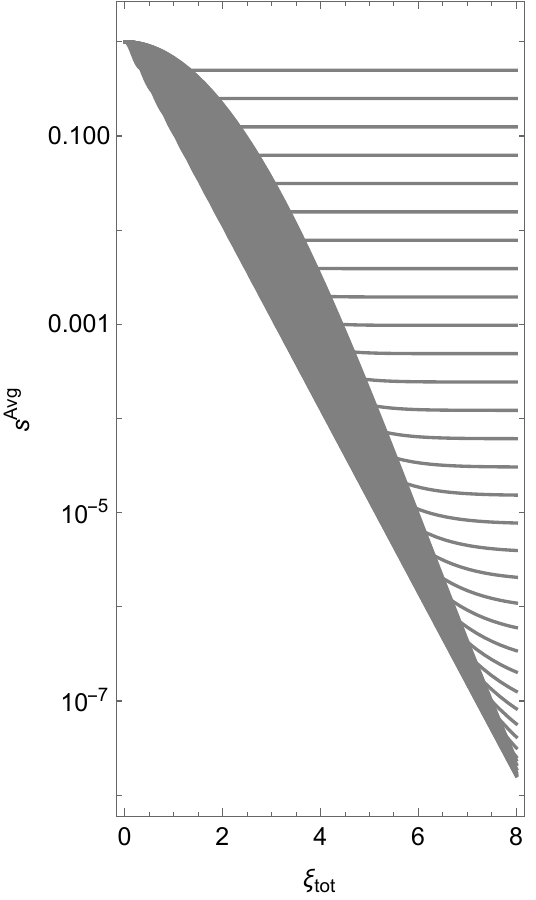}
    \includegraphics[width=0.33\textwidth]{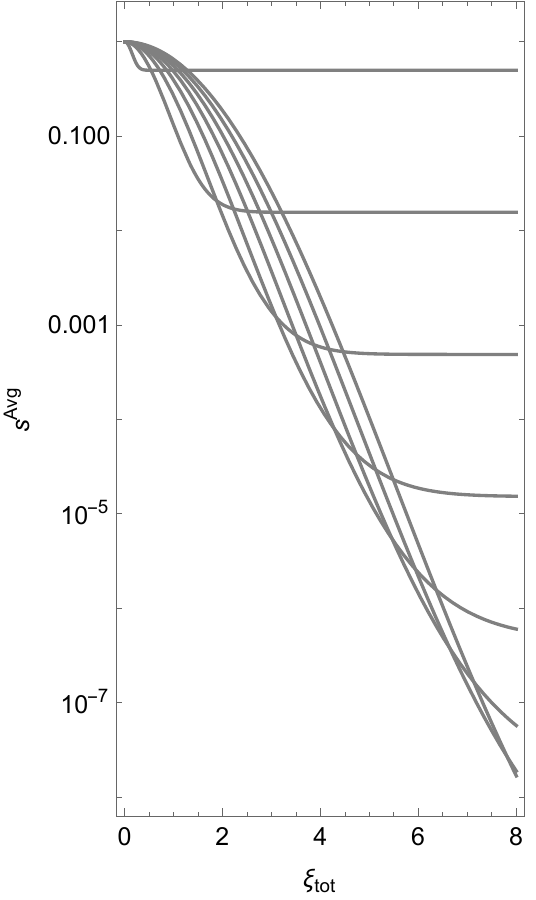}
    \includegraphics[width=0.33\textwidth]{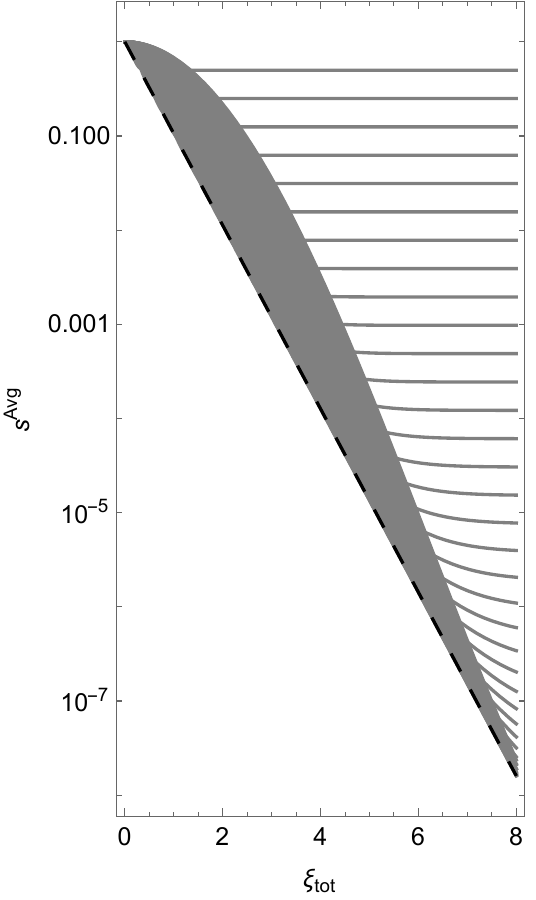}
    \caption{The average suppression factor as a function of $\zeta_{\rm tot}$ for various values of $n$.  The left panel shows all values of $n$ from 1 to 40. 
    As the curves become dense near the separatrix, obscuring the features, the middle panel shows every fifth value of $n$.  The right panel includes a dashed line corresponding to the exponential fit given in Eq.~(\ref{Eq:Mind}). }
    \label{fig:AVplot}
\end{figure*}

The original formulation of the algorithm chose times randomly from a Gaussian distribution with a predetermined standard deviation for each iteration~\cite{Choi:2020pdg}. We will refer to this incarnation of the rodeo algorithm as the random rodeo algorithm (RRA). The logic of RRA simple: there is no correlation between the times apart from that fixed by the overall standard deviation. Given this one might reasonably expect that all excited states will have large suppression after many iterations since one does not have coherent regions that are largely unsuppressed. 


This section will discuss how to optimize RRA to maximize suppression for minimal computational cost.  As will be discussed in subsequent sections, there exist other ways to pick times which generically produce more suppression at less computational cost than the optimized RRA.  But to see that, it is important to know just how well RRA can do.

The optimization problem is not yet well-posed.  Different components of the initial state are suppressed differently and depending on the information one has about the initial state one might sensibly alter what one is attempting to optimize.  For example, if one knew that the excited state components in the initial state was dominated almost entirely by components with energies between $\Delta$ (the minimum excitation energy)  and $2\Delta$, one would attempt to configure the algorithm to suppress those configurations optimally; the strategy could be qualitatively different if the initial state contained substantial contributions with excited energies covering many orders of magnitude.  

Let us consider the case where we know nothing about the initial state beyond the fact that it is normalized and all excited components have energy greater than or equal to $\Delta$. In that case, it seems reasonable to adopt a conservative strategy to optimize the suppression of the components with energies equal to or greater than $\Delta$ which on average are the least suppressed ({\it i.e.}, have the largest expectation value of the suppression factor $s$) assuming fixed average total time. Since this variant of the algorithm picks times randomly all one can specify is the average total time; similarly one cannot ensure that there will not be components that are less suppressed than average, but at least one can be sure that in a statistical sense the suppression should be at least as large as this for any component. Equivalently, one can ask how much time one needs to spend on average to ensure that the expected average suppression for all components with energy greater than $\Delta$ are suppressed on average by at least fixed predetermined factor.

This is straightforward to do. From Eq.~(\ref{Eq:suppress}), the suppression for a component with energy $E = \hbar \omega$ after $n$ iterations of the algorithm is 
$s(\omega) = \prod_{j=1}^{n} \cos^2 \left(\frac{\omega T_j}{2} \right)$ and since the $T_j$ are chosen randomly and independently of each other, the ensemble average of the suppression is given by
\begin{widetext}
\begin{equation}
\begin{split}
    s^{\rm Avg} &= \int \left(\prod_{j=1}^n \, dT_j \, \,  p(T_j, T)  \right)  \cos^2 \left(\frac{\omega T_1}{2} \right) \cos^2 \left(\frac{\omega T_2}{2} \right)\cdots \cos^2 \left(\frac{\omega T_n}{2} \right) =\left( \int d T' p(T', T) \cos^2 \left(\frac{\omega T'}{2} \right)\right)^n\\
    &= \left( \frac{1 }{2}\right )^n  \left ( 1+\exp \left(-\frac{\pi \omega^2 T^2}{4} \right) \right)^n = \left( \frac{1 }{2}\right )^n  \left ( 1+\exp \left(-\pi^3 \zeta^2  \right ) \right )^n 
\end{split} \label{Eq:AveSUp}\end{equation}
\end{widetext}
where $n$ is the number of iterations,  $p(T_j, T)$ is given from the positive half of a normal distribution whose average value is $T$ (so that $T= \sqrt{\frac{2}{\pi}} T_{\rm RMS}$) and the substitution $\zeta \equiv \frac{\omega T}{2 \pi}$ is made in the final form of $s$.  

It should be clear that as $\zeta \rightarrow 0$, $s^{\rm Avg} \rightarrow 1$.  This simply reflects the fact the algorithm is designed so that the ground state remains unsuppressed.  On the other hand, for large $\zeta$, $s^{\rm avg}$ asymptotes\footnote{Note this asymptotic behavior differs from the $1/4^n$ discussed in Ref.~\cite{Choi:2020pdg} that $1/4^n$ behavior corresponds to the geometric mean while the present result is for the arithmetic mean.  The qualitative difference between these behaviors reflects very large fluctuations and will be discussed in the next section } to $1/2^n$.  Moreover, it approaches its asymptotic value quite quickly since $\exp(-\pi^3 \zeta^2)$ becomes very small for quite modest values of $\zeta$.   

Note that the total average time, $T_{\rm tot}$, is simply the number of iterations times the average time per iteration given by $T_{\rm tot} = n T$.  It is useful to rewrite Eq.~(\ref{Eq:AveSUp}) in terms of $T_{\rm tot}$ or equivalently $\zeta_{\rm tot}=\omega T_{\rm tot} /2 \pi $: 
\begin{equation}  
s^{\rm Avg}=\left( \frac{1 }{2}\right )^n  \left ( 1+\exp \left(\frac{-\pi^3 \zeta_{\rm tot} ^2 }{n^2} \right ) \right )^n 
\label{Eq:AveSUp1}\end{equation}

In Fig.~\ref{fig:AVplot}, $s^{\rm Avg}$ is plotted as a function of $ \zeta_{\rm tot}$ for various values of $n$.  The leftmost panel includes all $n$'s from 1 to 40. It is clear that the $\zeta_{tot}-s^{\rm Avg}$ plane separates into two regions. Situations in which there are no values of $n$ for which a given value of $\zeta_{tot}$ has a corresponding $s^{\rm Avg}$ are in the lower left. In this region it is impossible to use RRA to obtain an average level of suppression, $s^{\rm Avg}$, for the  given $\zeta_{tot}$. Conversely, the top right region corresponds to situations in which it is possible to use RRA to obtain an average level of suppression, $s^{\rm Avg}$, for the  given $\zeta_{tot}$. There is a clear separatrix between these two regions.

The way the separatrix forms is a bit hard to discern from the left most panel of Fig.~\ref{fig:AVplot} since the curves for the various $n$'s are so dense. To clarify how the nearly exponential separatrix forms, the middle panel plots only every fifth $n$. For values of $\zeta_{tot}$ greater then 1, the separatrix is approximated extremely well by a simple exponential---a line in the log-plot.   

The most natural fit to a simple exponential is the one that exactly matches the exact separatrix once for each value of $n$. This is straightforward to do: if one takes the functional form of Eq.~(\ref{Eq:AveSUp1}) but takes $n$ to be a continuous real variable rather than having discrete integer values, one can minimize $S^{\rm Avg}$ with respect to $n$ for fixed $\zeta_{tot}$.
This yields:
\begin{subequations}
\begin{equation}
n = \alpha \zeta_{\rm tot} \label{Eq:Mina}
\end{equation}
where 
$\alpha$ satisfies
\begin{equation}
    \frac{2 \pi^2}{\alpha^2 \left(1+ \exp \left( \frac{\pi^3}{\alpha^2}\right) \right) } = \log \left (\frac{2}{1 + \exp \left(-\frac{\pi^3}{\alpha^2}\right) } \right)
\end{equation}
so that 
\begin{equation}
    \alpha  \approx 4.271 \label{Eq:Minc}
\end{equation}
and 
\begin{equation}\begin{split}
 & S^{\rm Avg}  = \exp \left (- \beta \,\zeta_{\rm tot} \right) \; {\rm with } \\ &\beta = - \alpha \log \left(\frac{1+\exp \left(-\frac{\pi^3}{\alpha^2}\right)}{2} \right) \approx 2.244 \; . \label{Eq:Mind}
 \end{split}
 \end{equation}
Equations (\ref{Eq:Mina})-(\ref{Eq:Mind}) give the minimum value of $S^{\rm Avg}$ for any value of $n$, integer or discrete. However, RRA is restricted to integer values. Thus, Eq.~(\ref{Eq:Mind}) gives a lower bound for $S^{\rm Avg}$ : 
\begin{equation}
\begin{split}
 & S^{\rm Avg}  \ge \exp \left (- \beta \, \zeta_{\rm tot} \right) \; {\rm with }\; \beta \approx 2.244 \; . \label{Eq:Mine}
\end{split}
\end{equation}
However, when $\zeta_{\rm tot}$ is such  that $n$ (as given in Eq.~(\ref{Eq:Mina})-(\ref{Eq:Minc})) happens to be an integer, then the simple exponential form precisely matches the separatrix and Eq.~(\ref{Eq:Mine}) becomes an equality.  Moreover, since the curves for the various $n$'s near the separatrix become quite dense by the time $\zeta_{\rm tot}$ reaches one, one expects the simple exponential form to be an excellent approximation.  This is seen in the right panel of Fig.~\ref{fig:AVplot}.
\end{subequations}

With this result in hand let us return to the optimization problem of determining how much time one needs to spend to ensure that all components with energy greater than $\Delta$ (the gap or minimum excitation energy) are suppressed on average by at least fixed predetermined factor $S$. Recalling that $\zeta_{\rm tot} = \frac{ E T_{\rm tot}}{2 \pi \hbar}$, Eq.~\ref{Eq:Mine} becomes 
$T_{\rm tot} \ge -\frac{2 \pi \hbar \log( S)}{ \, \beta E} $. To ensure that the time is long enough so that all components of the state are suppressed by at least $S$, one needs to use $\Delta$ for the energy.  Thus,
\begin{equation}
    T_{\rm tot} \ge -\frac{2 \pi \hbar \log( S)}{ \, \beta \Delta} 
\end{equation}
If one uses $T_0 \equiv 2 \pi \hbar/\Delta$ as a natural unit of time then
\begin{equation}
    \frac{T_{\rm tot}}{T_0} \ge -\frac{ \log( S)}{ \, \beta } \approx  .4456 \log \left( S \right ) . \label{Eq:beta}
\end{equation}

\section{Fluctuations in the random rodeo algorithm}\label{sec4}

Ensuring that the ensemble averaged suppression factor is smaller than some fixed value $S$ for all excited state energies might seem to be a sensible way to obtain a controlled level of suppression. This would certainly be true if the statistical fluctuations in the suppression factor were themselves under control. For example, if the suppression factors were distributed normally so that the ratio of the distributions width to the average suppression scaling like $n^{-\frac{1}{2}}$, then at large $n$ one could be quite confident that the suppression factor in various runs did not exceed $S$ much. This would allow a reliable and predictable projection onto the ground state.

Unfortunately, this is not the case. Rather, there are exceptionally large fluctuations, which make it very difficult (and computationally expensive) to produce a given level of suppression in a reliable way.

\subsection{Fluctuations in statistical measures over the random ensemble}

\begin{table*}[t!]
\centering
\begin{tabular}{|| c | c | c | c ||}
    \hline
    statistical  & $ s(\zeta_{\rm tot})$ for fixed total  & asymptotic form   &  best exponential fit for separatrix  \\
    quantity & average time and fixed $n$ & for large $\zeta_{\rm tot} $ & in the $\zeta_{\rm tot}-s$ plane\\
    \hline
    geometric mean & $\exp  \left(n \int_{-\infty}^\infty d T \log \left ( \cos^2 (\pi T \zeta_{\rm tot})\right) p(T, 1) \right)$ & $4^{-n}$ &
    $ \exp \left (- \beta \,\zeta_{\rm tot} \right) $ \; \; with \; \;  $\beta \approx 4.46 $ \\[2ex]
    \hline
    arithmetic mean & $\left( \frac{1}{2}\right )^n  \left ( 1+\exp \left(-\pi^3 \zeta_{\rm tot}^2 \right ) \right )^n$ & $2^{-n}$ & 
    $ \exp \left (- \beta \,\zeta_{\rm tot} \right) $ \; \; with \; \;  $\beta \approx 2.244$ \\[2ex]
    \hline
    root-mean-square & $\left( \frac{3}{8} \right)^{n/2 }  \left( 1 + \frac{\exp(-4 \pi^3 \zeta_{\rm tot}^2)}{3}  + \frac{4 \exp(-\pi^3 \zeta_{\rm tot}^2)}{3}  \right)^{n/2} $ & $\left(\frac{3}{8} \right)^{n/2 } $ &
    $ \exp \left (- \beta \,\zeta_{\rm tot} \right) $ \; \; with \; \;  $\beta \approx 1.637$ \\[2ex]
    \hline  
\end{tabular}
    \caption{Three statistical quantities characterizing the suppression factor as a function pf $\zeta_{\rm tot}$ given by the ensemble used in the random rodeo algorithm.  The second column gives an expression for the quantity as a function of  $\zeta_{\rm tot}$ for fixed $n$.  The third column gives the asymptotic value of these forms at large  $\zeta_{\rm tot}$.  In all cases the asymptotic form is reached to very good approximation whenever  $\zeta_{\rm tot} >1$. The fourth column gives the best exponential fit analogous to the separatrix in the $\zeta_{\rm tot}-s$ plane which were calculated analogously to the method described in the discussion surrounding Eqs.~\ref{Eq:Mina} - \ref{Eq:Mine}.  As there, the exponential fit to the separatrix determines the minimal time needed to ensure that the statistically weighted suppression factor achieves a fixed level or better for all energies.} \label{Table:T1}
\end{table*}

A hint that exceptionally large fluctuations are present  can be seen in the fact that there are multiple qualitatively different ``natural'' expectations for the amount of suppression after $n$ iterations. 

Since the average of $\cos^2$ is $\frac12$, it seems reasonable that well away from $\zeta=0$ (where there is no suppression) on average each iteration should randomly sample from $\cos^2$  leading to expectation of an average suppression of $\frac 12$ per iteration.  On the other hand, it also seems reasonable that well away from $\zeta=0$, where a random assignment of times will lead to an approximately uniform distribution of phases in the cosine and as noted in Ref.~\cite{Choi:2020pdg}, the geometric mean of $\cos^2(\theta)$ leads to an expectation of a suppression of $\frac 14$ per iteration.  


Which one of these ``natural'' expectations is correct? An apparently puzzling feature is that there is strong evidence that each of these is at least approximately correct: Ref.~\cite{Choi:2020pdg} gives evidence for a suppression of $\frac{1}{4}$ per iteration: it plots the logarithm of the suppression for 4 values of $\zeta$ for one random run of up to 150 and shows that for  $\zeta = 3.0$ the logarithm suppression closely follows $-N \log 4$ (where $N$ is the number of iterations (with notable fluctuations)  while for $\zeta = 2.0$ it nearly does so. This strongly supports the notion that the suppression per iteration is at least close to $\frac 14$ for sufficiently large $\zeta$.

However, the evidence that the (ensemble) average suppression per iteration is $\frac 12$ for large enough $\zeta$ is also compelling. From Eq.~(\ref{Eq:suppress}), the suppression for a component with energy $E = \hbar \omega$ for $n$ iterations of the algorithm is 
$s(\omega) = \prod_{j=1}^{n} \cos^2 \left(\frac{\omega T_j}{2} \right)$ and since the $T_j$ are chosen randomly and independently of each other, the ensemble average of the suppression is given by
Eq.~(\ref{Eq:AveSUp1}).  Clearly as $\zeta$ gets large, $s$, the ensemble average suppression, approaches $2^{-n}$, not $4^{-n}$.

These two asymptotic scaling behaviors may seem incompatible,  but they need not be. It is perfectly possible that ``typical'' suppression factors scale at least roughly as $4^{-n}$ for sufficiently large $\zeta$, while the average is dominated by atypical suppression factors that occur relatively infrequently but are orders of magnitude larger than the typical ones.  

It is easy to verify that this is happening.  In Table~\ref{Table:T1} the geometric mean and root-mean-square (RMS) values for the suppression for the ensemble used in RRA are given along with the arithmetic mean computed earlier. The expressions are derived analogously to the arithmetic mean.  

From the expressions for root-mean-square and arithmetic mean one can easily construct the ratio of the standard deviation $\sigma$ to the (arithmetic) mean $s^{\rm Avg}$: 
\begin{widetext}

\begin{equation}
    \begin{split}
\frac{\sigma}{s^{\rm Avg}}  =& \left(\left(\frac{3}{2} \right)^n \left(
\frac{  \left( 1 + \frac{\exp(-4 \pi^3 \zeta^2)}{3}  + \frac{4 \exp(-\pi^3 \zeta^2)}{3}  \right)}{  \left ( 1+\exp \left(-\pi^3 \zeta^2 \right ) \right )^{2} } \right)^{n} -1\right )^{\frac{1}{2}} 
\; \; \; \underrightarrow{{\rm large} \; n} \\ & \left(\frac{3}{2} \right)^\frac{n}{2} \left(
\frac{  \left( 1 + \frac{\exp(-4 \pi^3 \zeta^2)}{3}  + \frac{4 \exp(-\pi^3 \zeta^2)}{3}  \right)}{  \left ( 1+\exp \left(-\pi^3 \zeta^2 \right ) \right )^{2} } \right)^{\frac{n}{2}}  \; \; \; \underrightarrow{\rm large \; \zeta }\; \; \; \left( \frac{3}{2} \right)^{\frac{n}{2}} \; .
    \end{split}
\end{equation}
\end{widetext}
The behavior for large $n$ follows because the quantity
$$\left(\frac{3}{2} \right) \left(
\frac{  \left( 1 + \frac{\exp(-4 \pi^3 \zeta^2)}{3}  + \frac{4 \exp(-\pi^3 \zeta^2)}{3}  \right)}{  \left ( 1+\exp \left(-\pi^3 \zeta^2 \right ) \right )^{2} } \right)$$
is larger than unity for nonzero values of $\zeta_{\rm tot}$ and becomes much large than one for large $n$. Empirically, both the RMS and arithmetic mean approach their asymptotic quite rapidly at even modest values $\zeta_{\rm tot}$.

Clearly, at large $n$, the fluctuations in the distribution of $s$ values become exponentially larger as a function of $n$.  

The three statistical quantities which are all sample from the same distribution are different.  If the suppression factors were dominated by one characteristic scale, all of these quantities would be at the same scale. But as seen explicitly in the third column, three differ and the differences between them are exponentially large as a function of $n$.

\begin{figure}[t!]
    \centering
    \includegraphics[width=0.45\textwidth]{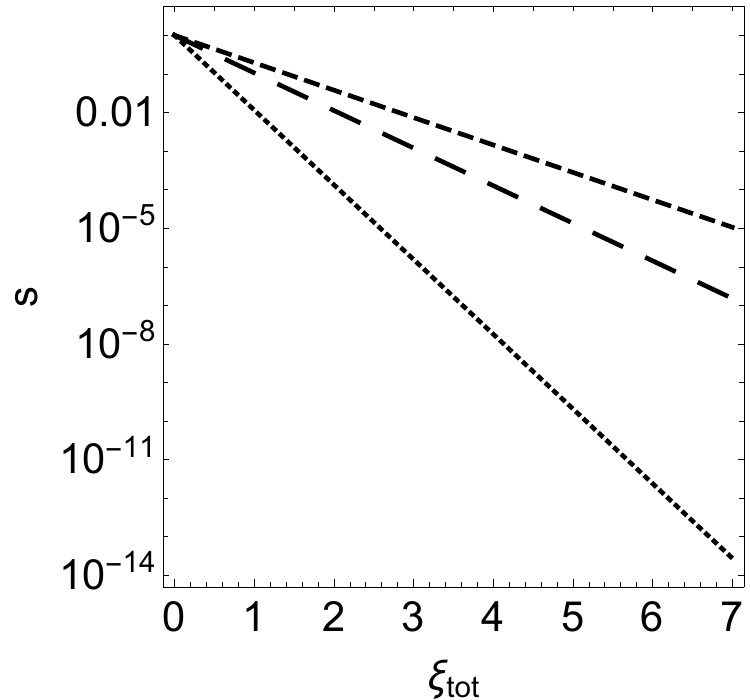} \;
    \caption{Exponential fits to  average (arithmetic mean) suppression factor (long dashed), root mean square suppression factor (short-dashed) and geometric mean (dotted) as a function of $\zeta_{\rm tot}$}
    \label{fig:Sepplot}
\end{figure}

To see how the large scale of the fluctuations affects the ability of the scheme to reliably accomplish a projection, the exponential fits in the fourth column of the Table~\ref{Table:T1} are shown in Fig.~\ref{fig:Sepplot}. The implication of this is quite striking.  If, for example, one considers the case of $\zeta_{tot} = 5$, RRA can be used to ensure that the geometric mean of suppression is below $2.07 \times 10^{-10}$, but the arithmetic mean of the suppression factor is at best $1.34 \times 10^{-5}$, nearly 5 orders of magnitude worse than the geometric mean; the best that can be achieved for the RMS is $2.79 \times 10^{-4}$ which is almost 6 orders of magnitude worse than the achievable value from the geometric mean.


It was suggested earlier that ``typical'' values of the suppression factors scale at least roughly as $4^{-n}$ for sufficiently large $\zeta$, explaining the results seen in Ref.~\cite{Choi:2020pdg}.  It is easy to understand why this occurs. The suppression factor is the product of many variables each of which is randomly chosen from  the same distribution. In such a situation one expects a log-normal distribution to emerge when the number of samples, $n$, becomes large: the central limit theorem can be applied to the logarithms of the variables which are combined  additively\footnote{Analogously with the central limit theorem for normal distributions, when the underlying distribution has a limited range, the range of the underlying distribution can put constraints on the ability of the tail of the log-normal distribution to correctly describe the true distribution for $n$ samples.  In the present context, the suppression factor is bounded from above by unity. However, the log-normal distribution has support for $S>1$. Thus, quantities which evaluated with the log-normal distribution receive significant contributions from regions of $S>1$ will not be correctly described by the log-normal distribution.  It turns out that the root-mean-square is such a quantity.}. The geometric mean of a log-normal distribution is known to be its median.  This implies that at large $n$ one expects that the median of the suppression factor---a reasonable proxy for a ``typical'' value---should be the geometric mean; in this case the geometric mean is easily shown to be  $4^{-n}$. 

\subsection{Fluctuations in a single run}

\begin{figure*}[t!]
    \centering
    \includegraphics[width=1.0\textwidth]{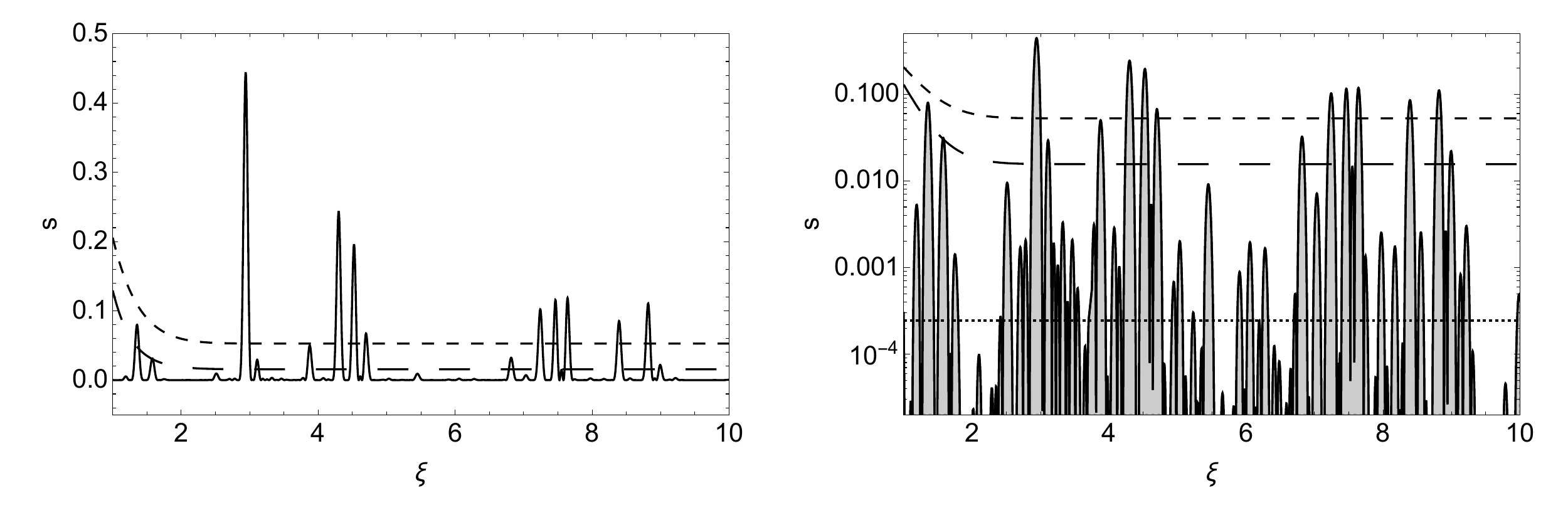}
    \caption{The solid line represents the suppression factor for a random run with six iterations with time chosen from the positive side of a normal distribution and an average time of unity.  The long-dashed curve is the ensemble average for 6 iterations with that distribution for the suppression, the short-dashed curve is the RMS and the dotted line in the log plot is the geometric mean.   }
    \label{fig:random}
\end{figure*}

So far, we have focused on the fluctuations of statistical quantities for fixed values of $\zeta_{\rm tot}$ evaluated over the statistical ensemble of RRA. However, intuitively, one might also expect that the large fluctuations seen in the ensemble average would be reflected as large fluctuations as a function of $\zeta$ for any given run. This subsection provides numerical evidence that this occurs by looking at a representative example. Indeed one sees that typical suppression factors are of order $4^{-n}$, the average suppression is of order $2^{-n}$, and the average suppression is dominated by narrow peaks in $\zeta$ with suppression factors that are much less suppressed than the average.

In Fig.~\ref{fig:random}, a representative sample is given for the suppression factor obtained after six iterations with times picked from the half-normal distribution with an average of unity and plotted over a range of $1 < \zeta < 10$. (Six-iteration was chosen so that the ensemble average  $1/64$ was still large enough to be easily discerned from zero by eye.) The existence of very large fluctuations compared to the average is readily apparent. The highest peak is 13 times larger than the average, while typical values between the peaks are so exceptionally small that they can only be discerned in a log plot.

The minimum value showed in the log plot gives an indication of just how small it is. It is clear that the fraction of value of $\zeta$ in the regime from 2 to 10 (where the average function has essentially asymptoted to $2^{-6}$) that has $s < 4^{-6}$ is close to half: a numerical integral of $\Theta(4^{-6}- s[\zeta])$ where $\Theta$ is the Heaviside step function yields an estimate that approximately 58\% of the points in this range have $s(\zeta ) <4^{-6}$ for this set of iterations. This is consistent with the notion that a ``typical'' point has $S(\zeta)$ of order $4^{-n}$. 

Similarly, the average value of $s$ in the region from 2 to 10 has an average value of $s(\zeta)$ of order $2^{-n}$. Numerically it is approximately 0.014 while $2^{-6} \approx 0.0156$, so for this particular run it is about 90\% of what one would expect the average to be, assuming that individuals reflect the ensemble average at large $n$ and that $n=6$ is sufficiently large. Given the paucity of large peaks in this range of $\zeta$ and the fact that only 6 iterations were used, this is remarkably close.

The existence of these large fluctuations represents a serious challenge for using RRA in an efficient way. In order to ensure that the suppression factor has a high probability of reaching a needed level of suppression for all relevant components of the initial state,  one must use computational resources that are several times larger than the times needed for a typical component to be suppressed to that level. If computational resources are fixed (as is likely when the first quantum computer becomes available for which the rodeo algorithm is viable for a particular problem) then the fluctuations imply that one does not even have a reliable estimate of the order of magnitude of the suppression, making one quite unsure of just how well an initial state has been projected onto the ground state.  

\section{Super iterations \label{Super}}

There are two reasons why non-random variants of rodeo projection might be preferable to RRA.  The first is related to the previous section: the random algorithm produces exponentially large fluctuations (in $n$) which greatly restrict the efficiency of the approach if one wishes to reliably know the level of suppression.  Ideally one could intentionally choose parameters that control these fluctuations. The second is simply that for any particular well-defined quantity associated with suppression there exists an optimal choice (in terms of $n$ and the various values of $\zeta$---or equivalently $T$---for each iteration) that maximizes (or minimizes as the case may be) the quantity. At best, a random algorithm can approximate this optimal choice.  Thus, as a matter of principle an appropriate intentional choice of the parameters should always perform better than the random algorithm---regardless of the quantity one is attempting to optimize.  

Note that large fluctuations arise when multiple iterations have suppression factors near unity. Suppression factors of unity occur when $\pi \zeta_c^j$ is an integer while suppression factors of zero occur when $\pi \zeta_c^j$ is an half-integer. One can restrict the scale of such large fluctuations by requiring that suppression factors of unity are always multiplied by suppression factors of zero for all non-zero values of $\zeta$. A key insight is that this can be arranged to occur with a finite total time (or $\zeta$) (and indeed only twice the time of the underlying distribution).

This can be done  using  ``super iterations'', rather than the single iterations considered heretofore. A super iteration consists of an infinite set\footnote{Of course in practice this will need to be large but finite; the effects of keeping it finite will be discussed in Subsection~\ref{subsec:cost}} of iterations each of which takes half the time (or $\zeta_c$) of the previous one. Since the times form a geometric series the net time for a super iteration is exactly twice of the underlying iteration.

Note from Eq.~(\ref{Eq:suppress}) that the second stage of the super iteration with a time given by $T^j/2$ has a suppression factor of zero whenever $\cos^2(\pi \zeta_c^j/2)$ is zero, which occurs when $\zeta_c^j$ is twice an half-integer (odd integer). Thus it coincides with places where the first stage has a suppression factor of unity, and exactly cancels them.  It does not cancel all of the place where the initial iteration was unity.  However the next stage---where the a time is given by $T^j/4$---will cancel when $\zeta_c^j$ is four times an half-integer, the stage after that when $\zeta_c^j$ is eight times an half-integer and so forth. Thus, as a whole the super iteration will cancel every place where original iteration was unity at a cost of spending twice the time~\footnote{It is analogous to the method suggested in~\cite{Stetcu:2022nhy} to remove states with discrete quantum numbers}.

The suppression factor for the $j^{\rm th}$ super iteration at given $\zeta_{c \, \, \rm sup}^j$ (which is twice $\zeta_c^j$, the base iteration,  since a super iteration takes twice as long) is given by 
\begin{equation}
\begin{split}
    s_{c \, \rm sup}^j &= \prod_{k=0}^{\infty} \cos^2 \left( \frac{\pi \zeta_{c}^j}{2^k} \right ) = \prod_{k=1}^{\infty} \cos^2 \left( \frac{\pi \zeta_{c \, \, \rm sup}^j}{2^k} \right )\\
    & = j_0^2 \left (\pi \zeta_{c \, \, \rm sup} ^j  \right ) = \left(\frac{\sin(\pi \,  \zeta_{c \, \rm sup})}{\pi \, \zeta_{c \, \, \rm sup}} \right )^2 \ \; 
\end{split}
\label{Eq:GeoS}
\end{equation}
where $j_0(x)$ is the $0^{\rm th}$ spherical Bessel function.

\begin{figure}[b!]
    \centering
\includegraphics[width=.45\textwidth]{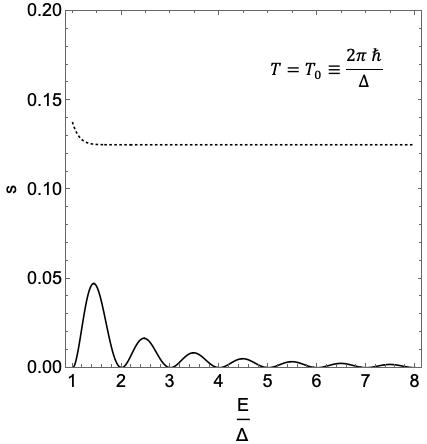} \;
    \caption{The solid line represents the suppression factor for a single ``super iteration''  as a function of the excitation  energy in units of $\Delta$, the minimum excitation energy.  The time was fixed to be $T_0 = 2 \pi \hbar /\Delta$. The dotted line uses the same total (average) time and gives the ensemble average suppression factor for three iterations---the number of iteration that ensures that the maximum value of the average suppression factor is as small as possible. }
    \label{fig:jo}
\end{figure}

From Eq.~(\ref{Eq:zeta}), $\zeta=(E/\Delta) (T/T_0)$.  Thus if one allocates a time of $T_0 = 2 \pi \hbar /\Delta$---one period of phase evolution for the lowest excited state---and does just a single super iteration, the suppression factor has a function of energy will be  given by $j_0 \left (\pi E/\Delta) \right )^2$ and is shown in Fig.~\ref{fig:jo}.

Fig. \ref{fig:jo}  plots $s$ for $E/\Delta \ge 1$, since  $\Delta$ is the minimum excitation so that all excited state components in the initial state have  $E \ge \Delta$. The choice of $T=T_0$ ensures that for the lowest excited state $s=0$, the components with a minimal excitation energy are completely suppressed. There are two other important salient features. The obvious one is evident from the scale of the suppression factor. Note that the maximum suppression factor is less than .05 for all energies above the lowest excited state. The other is that $s$ falls off rapidly with $E/\Delta$ or equivalently $\zeta$.

The overall small scale of $s$ for all relevant energy is quite striking given that the net time is very modest---only a single period of the lightest excitation. A useful comparison is with RRA: if one restricted the (average) total time of a single period of the lightest excitation, and optimized the average for $E>\Delta$, then one should use three iterations (see Fig.~\ref{fig:AVplot}). The average suppression factor is plotted as a dotted line. Note that for all values of $E/\Delta >1$ the super iteration is well below the average value of the random algorithm; it is always at a factor of 2.6 smaller. Moreover, for larger $E/\Delta >1$ the single super iteration rapidly becomes orders of magnitude better than three iterations of the random algorithm. The fact that super iterations drop off so rapidly at large $\zeta$ is a clear benefit in attempting to construct optimized versions of the rodeo algorithms---if a scheme using multiple super iterations concentrates on improving the suppression at values of $\zeta$ between one and two, it will be essentially guaranteed to have excellent suppression everywhere for $\zeta>1$. This is not true of the random algorithm.


\section{Optimizing rodeo projection}\label{sec6}

The demonstration that a single super iteration achieves substantially more suppression at all energies than the random algorithm with the same fixed time---and does so without suffering from large uncontrolled fluctuations---makes clear the advantages of intentionally choosing the parameters of the rodeo algorithm rather than relying on randomness to achieve optimal suppression.  

As noted in Section~\ref{OptRRA}, in order to optimize the rodeo projection algorithm one needs to know the quantity that one is optimizing. Even if one poses a problem as one fixes the total computational resources to be used (or more precisely its proxy, the total time in units of $T_0 =\frac{2 \pi \hbar}{\Delta}$) and asks to optimize the suppression factor, one has not fully specified the optimization problem since the suppression factor is a function of excited energies. 

\begin{table*}[t]
\centering
\begin{tabular}{||  m{.85in} |m{.5 in}|m{.65 in} |m{3.425 in} ||}
\hline

\centering maximum suppression factor & \centering total time in units of $T_0$ & \centering number of super iterations &  \centering times for each super iteration in units of $T_0$ \\  

\end{tabular} 
\renewcommand{\arraystretch}{1.6}
\begin{tabular}{|| m{.85 in} | m{.5in} | m{.65 in}|m{.375 in} | m{.375 in} | m{.375 in} | m{.375 in} |m{.375 in} | m{.375 in} | m{.375 in}  | m{.375 in}||}
\hline

\centering  $\underset{E \ge \Delta}{\rm max} \, \, \, s\left(\frac{E}{\Delta} \frac{T_{\rm tot}}{T_0} \right)$&\centering  ${T_{\rm tot}}/{T_0}$ & \centering $n$ & \centering ${T_1}/{T_0}$ & \centering ${T_2}/{T_0}$  & \centering  ${T_3}/{T_0}$ & \centering  ${T_4}/{T_0}$ &\centering  ${T_5}/{T_0}$ & \centering  ${T_6}/{T_0}$ & \centering ${T_7}/{T_0}$ & ${T_8}/{T_0}$ \\ 
\hline
\centering $4.719 \times 10^{-2}$ & \centering .8129 & \centering  1 & .8129& & & & & & & \\ \hline
\centering $8.508 \times 10^{-4}$ & \centering 1.5906 & \centering  2 & .9361& .6545& & & & & & \\  \hline
\centering $2.421 \times 10^{-5}$ & \centering 2.4222 & \centering  3 & .9494& .6638& .8090& & & & & \\ \hline
\centering $7.549 \times 10^{-7}$ & \centering 3.0752 & \centering  4 & .9785& .6841& .8338& .5788& & & & \\ \hline
\centering $7.385 \times 10^{-9}$ & \centering 3.9865 & \centering  5 & .9764& .6827& .8320& .5776& .9180& & &  \\ \hline
\centering $8.948 \times 10^{-11}$ & \centering 4.7944 & \centering  6 & .9881& .6908& .8419& .5845& .9290& .7601& &   \\ \hline
\centering $5.689 \times 10^{-12}$ & \centering 5.4500 & \centering  7 & .9925& .6939& .8457& .5871& .9331& .7634& .6343&   \\ \hline
\centering $1.539 \times 10^{-14}$ & \centering 6.4010 & \centering  8 & .9895& .6918& .8431& .5853& .9303& .7611& .6324& .9675         \\ \hline

\end{tabular}
\caption{The times for each super iteration and the maximum value of the suppression factors for $E>\Delta$ using the {\it ad hoc} scheme described in Appendix~\ref{App:WAM}.}\label{Table:T2}
\end{table*}

If one knew a significant amount about the spectral density of the initial state, then one might choose to optimize the algorithm in a manner which suppresses components that have high probabilities more heavily. Merely knowing information about the spectrum of the theory, without any addition information about the initial state could influence the choice of what we wish to optimize. For example, if one knew that the spectrum contained no states with energies between $\Delta$ and $2 \Delta$, then it would not make sense to expend computational resources suppressing components with energies in that region since one knows that they do not appear in the initial state.

The most straightforward approach to optimization has two steps: the first is to choose some quantity $Q(s,T_{\rm tot})$ that is a functional of the suppression factor  $s(\zeta_{\rm tot})$, and the total time,  and has the property that when minimized yields strong suppression given the state of interest.  The second step is to seek the $s$ produced by the rodeo algorithm subject to constraints on the total time that minimizes $Q$ as best one can.

Note that whatever strategy one adopts in choosing $Q$, there exists an optimal choice of parameters that minimizes $Q$ once $T_{\rm tot}$ is specified. That said, it may not be practical to find the optimal choice or even come close. One can explore a range of possibilities and ultimately use the one which provides the smallest value of $Q$.  

We start by considering the situation where nothing is known about the initial state beyond the fact that all unwanted components have energies greater than or equal to $\Delta$, and one wishes to adopt a conservative strategy in which one has a rigorous bound on the total suppression of excited states. There is a clear strategy: one can choose Q to be the largest value as a function of E in the physically relevant regime $E \ge \Delta$:
\begin{equation}
    Q=  \underset{E \ge \Delta}{\rm max} \, \, \, s\left(\frac{E}{\Delta} \frac{T_{\rm tot}}{T_0} \right), \label{Eq:Q}
\end{equation}
which, in effect, implies choosing the worst possible value of the suppression in the possible physical range. By choosing $Q$ to be the worst case, one ensures that the suppression factors for all components of the state are not larger than $Q$, which in turn puts an upper bound for $S_E$, and the total amount by which the excited state components are suppressed:
\begin{equation}
    S_E \equiv \frac{ \int_{\Delta-\epsilon }^{\infty} d E\, \,  s\left(\frac{E}{\Delta} \frac{T_{\rm tot}}{T_0} \right ) \rho(E)}{ \int_{\Delta-\epsilon }^{\infty} d E\, \,  \rho(E)} \le Q \; .\label{Eq:SEbound}
\end{equation}

\begin{figure}[b!]
    \centering
\includegraphics[width=.45\textwidth]{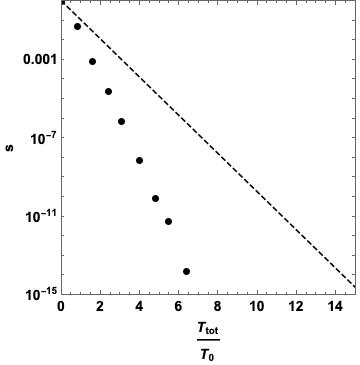} \;
    \caption{The solid points are an absolute upper bound on the suppression factor using intentionally fixed times the scheme outlined in Appendix~\ref{App:WAM} for various number of super iterations as a function of total time. For comparison, the dashed line represents a lower bound for the time for the ensemble average of $S$ to be greater than a fixed amount using the random rodeo algorithm of Ref.~\cite{Choi:2020pdg}.}
    \label{fig:comp}
\end{figure}

Moreover, it does so without using any information about the initial state or the spectrum (beyond the fact that lowest excitation is $\Delta$). Also, $Q$ depends solely on $s(\zeta_{\rm tot})$ and the total time, but not on the initial state. Thus one merely needs to seek to optimize $Q$ once for each $T_{\rm tot}$ and not undergo the daunting task of repeating the optimization for each initial state.

A potential downside of strategy (Eq.~(\ref{Eq:Q})) is its very conservative nature. As noted it will provide an upper bound for $S_E$, but it may be a very large upper bound compared to the least upper bound. Note that if we construct $s$ from super iterations with energies which are a few times $\Delta$, the suppression factor becomes extremely small near chosen energies even if only a few super iterations are used. If the initial state is heavily weighted towards such energies, the actual overall suppression of excited components, $S_E$, will be much smaller than the bound produced given by $Q$, and if the bound on $S_E$ is fixed at the level of projection needed for the problem at hand, substantially more computational resources would be expended than needed. After considering this conservative bound, we will consider ways to improve upon it.

Unfortunately, even for this most conservative approach, the mathematical problem of finding $s$ such that it yields the minimum value of  $\underset{E \ge \Delta}{\rm max} \, \, \, s\left(\frac{E}{\Delta} \frac{T_{\rm tot}}{T_0} \right)$  subject to the condition that $s$ is obtained through some number of rodeo iterations with fixed total time is an open one. However, one can adopt {\it ad hoc} prescriptions and test how well they do. It is natural to consider  prescriptions based on super iterations as these seem very efficient.

We illustrate the existence of prescriptions for which the times are chosen intentionally rather than randomly and that yield exceptionally small suppression factors in quite modest times using a scheme outlined in Appendix~\ref{App:WAM}. The prescription entails generating suppression factors based on a fixed number of super iterations of various times (all of order $T_0$), with each super iteration generated from the previous one. The times for each super iteration and the {\it maximum} suppression factors (for $E>\Delta$) are listed in Table~\ref{Table:T2}.  Note that the maximum suppression factors are remarkably small: for $T_{\rm tot}/T_0 \sim 3.$ the suppression factor is $\sim 10^{-6}$ while for $T_{\rm tot}/T_0 \sim 6.4$ the suppression factor is $\sim 1.5 \times 10^{-14}$.

It is instructive to compare the results using suppression factors obtained from the scheme in Appendix~\ref{App:WAM} with the random approach. In Fig.~\ref{fig:comp} the absolute {\it upper} bound of the suppression factor is compared with the {\it lower} bound for the time for the ensemble average of $S$ to be greater than a fixed amount. It is clear that for fixed total times the intentional approach gives suppression factors that are orders of magnitude better than the random approach, when $T$ is a couple of time $T_0$ or more. For example, when $T/T_0 =6.4$, the intentional approach yields an upper bound on the total suppression that is a factor of $3.77 \times 10^{7}$ smaller than the lower bound ensemble average. If the calculation is restricted to short times (which is likely to be true when quantum computers first become large enough to implement the rodeo algorithm with sufficient coherence times), the existence of achieving order of magnitude better performance at fixed times is striking.  

A less dramatic way to view the improvement is to focus on the total time needed to achieve a certain level of suppression, in which case the gain looks to be something like a factor of two in total time. Whichever way one views this, however, the comparison understates the advantages of using an intentional approach. As noted earlier, the random approach yields exponentially large fluctuations making it complicate the achievement with reliable levels of suppression. In contrast, this problem is entirely vitiated by choosing the times intentionally: it provides an upper bound for the suppression factor that is truly reliable.

Moreover, the scheme discussed in Appendix~\ref{App:WAM} was constructed entirely {\it ad hoc} to provide a concrete example of a scheme that substantially outperforms the random approach. It is not implausible that a broad search over parameter space could yield substantially improvements. We leave such a search to future research.  

As a practical matter, implementing rodeo projection efficiently would be helped greatly if one produces a table, similar to Table~\ref{Table:T2}, but containing a denser set of total times and ideally, better suppression factors. It would then be a simple matter to either pick out the suppression required for the problem at hand and find the time needed to ensure that this is achieved or alternatively fix a maximum time that can be allotted and find the level of suppression that could be achieved. In either case one could then use the times of various super iterations (or regular iterations) associated with that row of the table to implement the projection efficiently. 

\subsection{Incorporating information about the initial state}

If one had full information about the spectral function of the initial state, one could choose to optimize things differently. Instead of attempting to minimize  $Q=  \underset{E \ge \Delta}{\rm max} \, \, \, s\left(\frac{E}{\Delta} \frac{T}{T_0} \right)$ and thereby put an upper bound on $S_E$, the net suppression factor for the excited states, one could attempt to find a set of (super) iterations with fixed total time that minimized $S_E$ or a set with a fixed value of $S_E$ that minimized the time. This is unlikely to be done since one would need to do this minimization separately for each initial state whose spectral function is known. 

On the other hand, if one had access to a table of the sort described above (or even just Table~\ref{Table:T2}) and knowledge of the spectral function, one could either pick a time that one wishes to spend and find the precise value of $S_E$ associated with it (rather than merely having an upper bound that could be orders of magnitude larger than $S_E$) or perhaps more practically identify the level of suppression in $S_E$ required for the problem at hand and choose the shortest time from the table that achieves it. This time will likely be significantly shorter than the time obtained from the upper bound on $S_E$.


\begin{figure}[t!]
    \centering
\includegraphics[width=.45\textwidth]{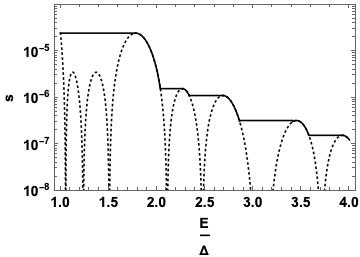} \;
    \caption{The solid line is $s^{\rm UB}\left(\frac{E}{\Delta} \frac{T}{T_0} \right)$, the least upper bound of $s\left(\frac{E}{\Delta} \frac{T}{T_0} \right)$ subject to the constraint that it is monotonically decreasing for the n=3 case given in Table~\ref{Table:T2}; the dotted line is $s\left(\frac{E}{\Delta} \frac{T}{T_0} \right)$ itself.}
    \label{fig:UB3}
\end{figure}

However, circumstances in which one has access to the full information of spectral function  will presumably be quite rare.  It seems far more likely that partial information about the spectral function might be available. Even with partial information one may be able to find significantly stronger upper bounds than Eq.~(\ref{Eq:Q}). 

To do this it is useful to construct a function that is the least upper bound of $s\left(\frac{E}{\Delta} \frac{T}{T_0} \right)$ subject to the constraint that it is monotonically decreasing. Let us denote this function  $s^{\rm UB}\left(\frac{E}{\Delta} \frac{T}{T_0} \right)$. In Fig.~\ref{fig:UB3}, this is done for the $n=3$ case given in Table~\ref{Table:T2}.   

Next, suppose that there is a good reason to believe that the most of the spectral strength above the ground state is predominantly at energies at least a few times $\Delta$. One can characterize this with an energy $E_0$ and $f$, the fraction of the excited state integrated spectral density above $E_0$: 
\begin{equation}
    f\equiv \frac{ \int_{E_0}^{\infty} d E\, \,   \rho(E)}{ \int_{\Delta-\epsilon}^{\infty}   dE \,  \rho(E)}\; .
\end{equation}
The bound on $S_E$ then becomes 
\begin{equation}
\begin{split}
    S_E & \le (1-f) \, s^{\rm UB}\left(\frac{\Delta}{\Delta} \frac{T_{tot}}{T_0} \right) +f  \, s^{\rm UB}\left(\frac{E_0}{\Delta} \frac{T_{tot}}{T_0} \right) \\ &\le Q \, .
\end{split}
\end{equation}
As an example, if one uses 3 super iterations with the simple scheme of Appendix~\ref{App:WAM} given in Table \ref{Table:T2}, and has $f=.99$ and $E_0=3$, then the bound becomes $S_E \le 5.591  \times 10^{-7}$, which is more than a factor of 40 smaller than the bound in Table~\ref{Table:T2} which assumes no knowledge of the initial state. If $f=.9999$ with $E_0=8$ the bound becomes $S_E \le 1.194 \times 10^{-8}$, which is a factor of 2000 smaller. 


\section{Practical considerations}\label{sec7}

A number of practical issues involving the implementation of the rodeo projection algorithm will arise when suitable quantum computers become available. This section will consider a few of these and discuss ways to deal with them.

\subsection{Cost of super iterations}\label{subsec:cost}

In the introduction it was suggested that the time of computation in units of $T_0$ is a reasonable proxy for the computational cost. While this is largely true, there is an important caveat, apart from allowing the system to evolve in time according to Hamiltonian dynamics, there is an additional overall computational cost associated with running each iteration: one needs to both couple the auxiliary qubit to the system of interest (which involves use of Hadamard gates and a conditional coupling) and to measure the auxiliary qubit. While such costs may be negligible for any given iteration, super iterations as formally defined in Section~\ref{Super} require an infinite number of iterations and thus, these fixed costs per iteration would become arbitrarily large.

The obvious solution is to simply use a fairly large but finite number of iterations for each super iteration. If one uses $N$ iterations in each super iteration, then 
\begin{equation}
s_{ \rm sup \, N}(\zeta_{\rm sup}) = 
\prod_{k=1}^{N} \cos^2 \left( \frac{\pi \zeta_{\rm sup}^j}{2^k} \right ) = \frac{j_0^2 \left (\pi \zeta_{\rm sup} \right ) }{j_0^2 \left (\frac{\pi \zeta_{\rm sup}}{2^{N}}  \right )}
\label{Eq:GeoS1} \; .
\end{equation}
Since $j_0^2 \left( \frac{\pi \zeta_{\rm sup}}{2^{N}} \right)$ is very close to unity until $\zeta_{\rm sup}$ is of order $2^{N-2}$, such a limitation will not noticeably alter any results for suppression until $\zeta_{\rm sup}$ gets very large.  

One might worry that when $\zeta_{\rm sup}$ is an integer multiple of $2^{N}$ the super iteration becomes unity and suppression is lost. This is unlikely to be a practical problem but it does mean that theoretical upper bound for any given $N$, the suppression factors given in a table analogous to Table \ref{Table:T2} (or a more sophisticated table) only hold for energies less than some calculable maximum. Thus, if one knows the integrated spectral density above that maximum is negligible, the bound remains valid. Fortunately such maxima occur at very high energies for relatively modest $N$. For example, when a single super iteration is used, the maximum energy (in units of $\Delta$) consistent with the suppression on the table will occur at approximately $2^N$. With the values in Table~\ref{Table:T2} a more precise maximum energy consistent with the quoted bound is $E_{\rm max}/\Delta \approx (2^{N}-1.430)/.8129$ so that for $N=15$, it holds up to energies of $40, 308 \times \Delta$. If more than one super iteration is used, the maximum suppression factor calculated for an arbitrarily large number of iterations per super iterations will continue to hold to much higher energies than the case of a single iteration. In any case, one can always make $N$ sufficiently large that this is not a practical problem.

\subsection{False negatives}

Up until now we have focused on reducing the probability of ``false positives'', the survival of components other than the ground state after the algorithm has run. We have ignored the problem of ``false negatives'' since, in the ideal case where the ground state energy is known precisely and the time evolution and control over timing are perfect, every ground state component will survive each iteration with unit probability. However, in practice the energy of the ground state is never known exactly and there will be errors due to imperfections in the propagation of the state in time. These can lead to ``false negatives'' in which the auxiliary qubit is measured in the down state even though the physical system is in the ground state. If one has a reasonable idea of the scale of the uncertainty of the ground state, the effects of imperfect calculation, and the probability that the state is in its ground state, the likelihood of a false negative for any given iteration can be estimated. Assuming that likelihood is small but not totally negligible, the natural thing to do is ignore the fact that the auxiliary qubit was measured down and continue the calculation while keeping track of a nominally unsuccessful run. If the auxiliary qubit is measured to be up through the next super iterations one should simply assume that a false negative occurred and continue. If a second nominally unsuccessful iteration occurs anytime up through the next super iteration one should assume that the iteration was truly unsuccessful. It should be very rare for this to care accidentally unless the uncertainty in the ground state energy is very large.

\subsection{Uncertainties in the energy of the first excited state}

The formulation of the problem assumed that the energy of the ground state and the first excited state were known. As discussed above a lack of precise knowledge of the ground state energy can lead to false negatives, but there is a useful strategy for dealing with these. A lack of precise knowledge of $\Delta$, the energy of the first excited state relative to the ground state, is even easier to mitigate. There are two possibilities to consider: either are the estimate of $\Delta$ is larger than the true value, or it is smaller. 

Underestimating $\Delta$ has a very modest cost. $T_0$, the time standard, will be larger than needed and hence all times (the proxy for computational cost) will be slightly larger than strictly necessary to achieve the level of suppression intended. However, the upper bound for suppression will be computed in Table~\ref{Table:T2} or its generalization will remain valid since a given value of $\zeta_{\rm tot}$ occurs at a lower energy.

Overestimating $\Delta$ is more serious: a given value of $\zeta_{\rm tot}$ occurs at a lower energies and hence effectively probes the suppression factor at values of the energy below the true value $\Delta$ where the suppression factor can be and generically is larger than the computed upper bound.

The obvious way to make sure that $\Delta$ is not overestimated is to ensure that the value used is chosen conservatively to be below the estimated value by an amount larger than the uncertainty. This would guarantee a reliable upper bound and a very modest computational cost.



\begin{acknowledgments}

This work was supported in part by the U.S. Department of Energy, Office of Nuclear Physics under Award Number(s) DE-SC0021143, and DE-FG02-93ER40762.

\end{acknowledgments}

\bibliography{refs.bib}

\newpage

\appendix
\section
{A rodeo iteration \label{sec:it}}
This appendix derives various properties of the density matrix and reduced density matrix associated with a single iteration of the rodeo algorithm. 

The Hilbert space of the total system is enlarged by the inclusion of an ancilla qubit, which is initialized to be in the state $|\uparrow \rangle$ (in the $z$ basis) and the physical system is in a pure quantum state.  Given the inclusion of an additional degree of freedom---which will be entangled with the state of the physical state, a natural description of the system the system is via density matrices and reduced density matrices.

The physical system prior to the running of the $j^{\rm th}$ iteration of the algorithm is $|\psi \rangle_{\rm phys}^{j \, \rm initial}$ given in Eq.~(\ref{Eq:initialstate});  the full (pure state) density matrix in the extended Hilbert space is  given by the tensor product  $$\hat{\rho}^{j \; \rm initial}_{\rm total}= |\uparrow\rangle \langle \uparrow| \otimes \hat{\rho}_{\rm phys}^{j \, \rm initial} \; ,$$ where the subscripts``phys'' and ``total'' indicate the space describing the physical system being studied and the total system including the ancilla qubit; $j$ simply labels the iteration.  

The total system is then subject to  unitary evolution:
\begin{equation}
\begin{split}
&\hat{\rho}^{j \; \rm final}_{\rm total}   = \hat{U}_j \, \hat{\rho}^{j \;\rm initial}_{\rm total}  \,\hat{U}_j^\dagger \; \; {\rm where} \\
& \hat{U}_j  =\exp \left (-i \left (\hat{p}^{\uparrow}_x \otimes \hat{H}_{\rm phys} T_j \right ) \right )  \; \; {\rm with }\\
&\hat{p}^{\uparrow}_x \equiv  \frac{1}{2} (1 + \hat{\sigma}_x) \; ;
\label{Eq:A1}\end{split}
\end{equation}
$\hat{p}^{\uparrow}_x$ acts on the ancilla qubit and is a projector onto the up state in the $x$ basis.  This is achieved by acting on the ancilla qubit with a Hadamard gate, running Hamiltonian dynamics for the physical system for a time $T_j$, conditionally on the ancilla qubit being in the ``up'' state and then again acting on the ancilla qubit with a Hadamard gate.  

The crux of the algorithm is the quantum interference between the components in which the Hamiltonian time evolves and the components where it does not. 


Following the time evolution the density matrix for the full system is given by 
\begin{widetext}

\begin{equation}
    \begin{aligned}
    & \hat{\rho}^{j \; \rm final }_{\rm total}  = |\uparrow\rangle \langle\uparrow| \otimes \biggl( |\alpha_g^j|^2 |\psi_g\rangle\langle \psi_g| 
    + \sum_{c , c'} \cos\left(\frac{\omega_c T_j}{2} \right)  \cos\left(\frac{\omega_c' T_j}{2} \right)   \left ( \alpha_c \alpha_c^* \,  |\psi_c\rangle\langle \psi_{c'}| + \alpha^*_c \alpha_{c'} \,  |\psi_{c'}\rangle\langle \psi_c| \right ) \\
    & + \sum_c \cos\left(\frac{\omega_c T_j}{2} \right)  \left( \alpha_g \alpha_c^* e^{-i \omega_c T_j / 2} \,  |\psi_g \rangle\langle \psi_c|
    + \alpha^*_g \alpha_c e^{i \omega_c T_j / 2} \,  |\psi_c\rangle\langle \psi_g|  \right) \biggr)   \\
    & + |\downarrow\rangle \langle\downarrow| \! \otimes \! \left( \sum_c |\alpha_c^j|^2 \sin^2\left(\frac{\omega_c T_j}{2} \right) |\psi_c\rangle \langle \psi_c|  + \sum_{c \ne c' } \sin\left(\frac{\omega_c T_j}{2} \right)\sin\left(\frac{\omega_{c'} T_j}{2} \right) \left ( \alpha_c \alpha_c^* \,  |\psi_c\rangle\langle \psi_{c'}| 
    + \alpha^*_c \alpha_{c'} \,  |\psi_{c'}\rangle\langle \psi_c| \right )\right) \\
    & + |\downarrow\rangle \langle\uparrow| \otimes \left( \sum_c \frac{1}{2} \left(e^{-i \omega_c T_j} -1 \right) \alpha_c | \psi_c \rangle \right) \left( \alpha_g^* \langle \psi_g | + \sum_{c'} \frac{1}{2} \left( e^{i \omega_{c'} T_j} + 1 \right) \alpha_{c'}^*  \langle \psi_{c'} | \right) \\
    &  + |\uparrow\rangle \langle\downarrow| \otimes \left( \alpha_g^* | \psi_g \rangle + \sum_{c} \frac{1}{2} \left( e^{-i \omega_{c} T_j} + 1 \right) \alpha_{c}^*  | \psi_{c'} \rangle \right) \left( \sum_c \frac{1}{2} \left(e^{i \omega_{c'} T_j} -1 \right) \alpha^*_{c'} \langle \psi_c | \right) \; {\rm with} \; \omega_a =\frac{E_a}{\hbar}.
    \label{Eq:density1}
    \end{aligned}
\end{equation}
The next step is to measure the state of the auxiliary qubit. This allows one to  use Eq.(\ref{Eq:density1}) to write the reduced density matrix  for the physical system at the end of the iteration in the following form:
\begin{subequations}
\begin{align}
  \hat{\rho}^{j \; \rm final }_{\rm phys} &  = |\alpha_g^j|^2 |\psi_g\rangle\langle \psi_g|  + \sum_c |\alpha_c^j|^2  |\psi_c \rangle \langle \psi_c| +  
\sum_c \cos\left(\frac{\omega_c T_j}{2} \right)  \left ( \alpha_g \alpha_c^* e^{-i \omega_c T_j / 2} \,  |\psi_g\rangle\langle \psi_c| 
+ \alpha^*_g \alpha_c e^{i \omega_c T_j / 2} \,  |\psi_c\rangle\langle \psi_g| \right ) \label{Eq:redden} \\  
& + \sum_{c \ne c'} \left ( \cos\left(\frac{\omega_c  T_j}{2} \right)  \cos\left(\frac{\omega_c' T_j}{2} \right)  
+ \sin\left(\frac{\omega_c T_j}{2} \right)  \sin\left(\frac{\omega_c' T_j}{2} \right) \right )  
\left ( \alpha_c \alpha_c^* \,  |\psi_c\rangle\langle \psi_{c'}| + \alpha^*_c \alpha_{c'} \,  |\psi_{c'}\rangle\langle \psi_c| \right ) \nonumber\\ \nonumber \\
  & =  P^{j \, \rm s}  \hat{\rho}^{j \, s} +   P^{ j\, u}    \hat{\rho}^{j \, \rm u} \;  \; 
{\rm with}  \label{Eq:rewrite}  \\
P^{j \, s} &  = |\alpha_g|^2 + \sum_c |\alpha_c^j|^2 \, \cos^2\left(\frac{\omega_c 
T_j}{2} \right) \; \;  ,  \; \; 
      P^{u \, {\rm s} }  = \sum_c |\alpha_c^j|^2 \, \sin^2\left(\frac{\omega_c 
T_j}{2} \right) \\ 
\hat{\rho}^{j \, s} &= \frac{\left ( \alpha_g |\psi_g\rangle + \sum_c \alpha_c^j \frac{1}{2} \left( 1 + e^{-i \omega_c T_j}  \right) |\psi_c\rangle \right ) \left ( \alpha_g^* \langle \psi_g| + \sum_c' \alpha_{c'}^{* j}  \frac{1}{2} \left( 1 + e^{i \omega_{c'} T_j}  \right) |\langle \psi_{c'}| \right )}{ |\alpha_g^j|^2   + \sum_c |\alpha_c^j|^2 \cos^2\left(\frac{\omega_c 
T_j}{2} \right)  } \\ 
 \hat{\rho}^{j \, u} & = \frac{\left ( \sum_c \alpha_c^j \sin\left(\frac{\omega_c T_j}{2} \right) |\psi_c\rangle \right ) 
 \left ( \sum_{c'} \alpha_{c'}^{* j}  \sin\left(\frac{\omega_{c'} 
T_j}{2} \right) |\langle \psi_{c'}| \right )}{  \sum_c |\alpha_c^j|^2 \sin^2\left(\frac{\omega_c 
T_j}{2} \right)  } \; ;
\label{Eq:iteffets}
\end{align}
\end{subequations}
\end{widetext}
in  Eq.~(\ref{Eq:rewrite}), $P^{j \, s}$ is the probability that the auxiliary qubit is in the up state (in the $z$) and  $P^{j \, u}$ is the probability that it is in the down state;   the superscripts, s and u denote ``successful'' and ``unsuccessful'' iterations.  

The nomenclature ``successful'' and ``unsuccessful'' are associated with the measurement of the auxiliary.  If it is measured  to be down  (in the $z$ basis), the entanglement with the auxiliary qubit ensures that the physical system is projected onto the state described by $\hat{\rho}^{j \, u}$ which has no components in the ground state.   Since the goal of the algorithm is to enhance the components in the ground state, such a measurement is regarded as an unsuccessful iteration. If the measurement yields an up state---a successful iteration---then the physical system is projected onto the state described by $\hat{\rho}^{j \, u}$. This is the density matrix for a pure state: the effect of a successful iteration is to transform the state of the physical system
\begin{equation}
\begin{split}
|\psi\rangle_{\rm phys}^{j \, \rm initial} &\xrightarrow[{\rm iteration}]{{\rm successful}} |\psi\rangle_{\rm phys}^{j \,  \rm final} \,\, {\rm with} \\
|\psi \rangle_{\rm phys}^{j \, \rm initial} & = \alpha_g |\psi_g \rangle + \sum_c  \alpha_c|\psi_c \rangle \,\, {\rm and} \\
|\psi \rangle_{\rm phys}^{j \, \rm final} &= \frac{ \alpha_g |\psi_g \rangle + \sum_c \frac{1}{2} \left( 1 + e^{-i \omega_c T_j}  \right) \alpha_c^j |\psi_c\rangle}{\sqrt{ |\alpha_g^j|^2   + \sum_c |\alpha_c^j|^2 \cos^2\left(\frac{\omega_c T_j}{2} \right) } } \; .
\end{split}
\end{equation}

\section{An {\it ad hoc} prescription \label{App:WAM}}

Table \ref{Table:T2} was obtained using a simple {\it ad hoc} prescription. The purpose was simply to demonstrate explicitly that there exist schemes that significantly outperform the average results of RRA and they do so without suffering from the exponentially large fluctuations which are inherent in the RRA scheme. It should be noted, however that there is no reason to assume, {\it a priori}, that this prescription achieves something close to the actual optimum choice.

We denote this, the Whac-a-Mole (WAM) prescription after the arcade game of that name, which bears a certain resemblance to the prescription. In that game, ``moles'' pop up in random places and need to be suppressed by whacking them with a soft mallet after which a mole emerges somewhere else, which also needs to be suppressed via whacking.  

The prescription is based on super iterations. As a first stage, one chooses a super iteration with a time  equal to $T_0$. This yields a suppression factor with zeros at all integer values of $E/\Delta$.  Our interest is in energies greater than $\Delta$. One can identify the largest value of the suppression factor for energies greater than $\Delta$ and consider this to be the ``mole'' to be whacked by choosing the time for the next supper iteration to create a zero a precisely at the energy of the maximum. Such a time will be larger than $T_0$ so that smallest zero in the suppression function will remain at $E=\Delta$. To do a third iteration one finds the maximum suppression factor for $E>\Delta$ and treats it as the ``mole'' to be whacked by choosing the time for the next super iteration to create a zero a precisely at the energy of the maximum. This can be continued for as many cycles as one wishes. Each cycle of this prescription yields progressively narrower peaks at progressively lower scales of the suppression factor and the maxima appear for $E/\Delta$ between one and two.  

The scheme as described so far is constrained to preserve $E=\Delta$ as the smallest zero of the suppression factor as a function of energy. Such a constraint can be relaxed to further optimize the time by simply using an overall scaling factor of less than unity for the times. Such a scaling does not affect the maximum value of the suppression factors but merely pushes a given value to higher energies. In other words, it acts to push energies that were previously below $\Delta$ to above $\Delta$. Since the suppression factor at $E=\Delta$ was initially zero, an overall scaling factor in times that is slightly smaller than unity will only push small suppression factors into the physical region, $E \ge \Delta$. To optimize the time, one chooses a scaling factor such that the new suppression factor at $E=\Delta$ is equal to the maximum, since any smaller one will increase the maximum. This scaling provides a very modest reduction in time after the first couple of cycles, but it is a nearly 20\% effect if one is using a single super iteration.

\end{document}